\documentclass[preprint,preprintnumbers,amsmath,amssymb]{revtex4}
\usepackage{amsfonts}    
\usepackage{amssymb}
\usepackage{latexsym}
\usepackage{eepic}
\usepackage{graphicx}
\usepackage{subfloat}
\usepackage{subfig}
\usepackage{multirow}
\usepackage[active]{srcltx}
\newcommand{\br}{\rule[0.5cm]{0.001cm}{0.0cm}}  

\newcommand{\al}{\alpha}
\newcommand{\pa}{\partial}

\newcommand{\La}{\Lambda}
\newcommand{\si}{\sigma}
\newcommand{\la}{\lambda}

\newcommand{\de}{\delta}

\newcommand{\De}{\Delta}

\newcommand{\rar}{\rightarrow}
\newcommand{\lrar}{\leftrightarrow}

\begin{document}

\title{The H$_2^+$ molecular ion: a solution}

\author{A.V.~Turbiner}
\email{turbiner@nucleares.unam.mx}
\author{H.~Olivares-Pil\'on}
\email{horop@nucleares.unam.mx}
\affiliation{Instituto de Ciencias Nucleares, Universidad Nacional
Aut\'onoma de M\'exico, Apartado Postal 70-543, 04510 M\'exico,
D.F., Mexico}


\begin{abstract}
Combining the WKB expansion at large distances and Perturbation Theory at small distances it is constructed a compact uniform approximation for eigenfunctions.
For lowest states $1s\si_{g}$ and $2p\si_{u}$ this approximation provides the relative accuracy $\lesssim 10^{-5}$ (5 s.d.) for any real $x$ in eigenfunctions and for total energy $E(R)$ it gives 10-11 s.d. for internuclear distances $R \in [0,50]$. Corrections to proposed approximations are evaluated. Separation constants and the oscillator strength for the transition $1s\si_{g} \rar 2p\si_{u}$ are calculated and compared with existing data.
\end{abstract}


\maketitle

\centerline{INTRODUCTION}

\vskip 1cm

The H$_2^+$ molecular ion is the simplest molecular system which exists in Nature.
It was the first studied molecular system since the inception
of the quantum mechanics which later appears in all QM textbooks (see e.g. \cite{LL}). Needless to say that this system plays very important role in different physical sciences, in particular, in laser and plasma physics.

From technical point of view this is the unique molecular system which admits complete separation of variables (in elliptic coordinates). Definitely, this problem  is non-solvable. Thus, the problem can be solved in approximate way only. We introduce a natural definition of solvability of non-solvable spectral problem: for any eigenfunction $\Psi$ we can indicate constructively an uniform approximation $\Psi_{app}$ such that
\begin{equation}
\label{delta}
    \vert \frac{\Psi(x) - \Psi_{app}(x)}{\Psi(x)}  \vert \lesssim \de \ ,
\end{equation}
in the coordinate space. It implies that any observable, any matrix element can be found with accuracy not less than $\de$. A simple idea we are going to employ is to combine WKB-expansion at large distances with perturbation theory at small distances near extremum the potential in one interpolation. Recently, this idea was realized for quartic anharmonic oscillator \cite{Turbiner:2005} and double-well potential \cite{Turbiner:2010}. In both cases for the lowest states it was constructed two-three parametric uniform approximations of the eigenfunction leading to 10 s.d. in energies and with $\de \sim 10^{-5}$ for any value of the coupling constant and size of the barrier. The goal of this Letter is to present such an approximation with $\de \sim 10^{-5}$ for two lowest (and the most important) states $1s\si_{g}$ and $2p\si_{u}$ of the H$^+_2$ molecular ion. It is worth mentioning that a study of the wavefunctions of the H$^+_2$ molecular ion in a form of expansion in some basis was initiated by Hylleraas \cite{Hylleraas:1931} and was successfully realized in the remarkable paper \cite{Bates:1953} (see also \cite{Montgomery:1977, Bishop:1978}). Attempts to find bases leading to fast convergence are still continuing. At present, the pure exponential basis seems the most fast convergent (see e.g. \cite{Korobov:2000} and references therein). It is worth noting that following the analysis of classical mechanics of the H$_2^+$ system and its subsequent semiclassical quantization some uniform approximations of wavefunctions of low lying electronic states were constructed \cite{Strand:1979}. Local accuracies of these approximations are unclear albeit eigenparameters are found with a few significant figures.

The Schr\"odinger equation, which describes the electron in the field of two centers of the charge $Z=1$ at the distance $R$, is of the form
\begin{equation}
\label{Sch}
    \left(-\De - \frac{2}{r_1}- \frac{2}{r_2}\right)\Psi \ =\ E' \Psi\ ,\
    \Psi \in L^2 ({\bf R^3})\ ,
\end{equation}
where $E'=(E - \frac{2}{R})$ and the total energy $E$ are in Rydbergs, $r_{1,2}$ are the distances from electron to first (second) center, respectively. Following \cite{LL} let us introduce the dimensionless elliptic coordinates \footnote{From $3D$ point of view they are {\it prolate spheroidal}.}:
\begin{equation}
\label{ell}
  \xi = \frac{r_1+r_2}{R}\ ,\quad \eta = \frac{r_2-r_1}{R}\ ,\quad 1 \leq \xi \leq \infty\ ,\quad -1 \leq \eta \leq 1
\end{equation}
and azimuthal angle $\phi$. The Jacobian is $\propto (\xi^2-\eta^2)$. The equation (\ref{Sch}) admits separation of variables in (\ref{ell}). Since the projection of the angular momentum to the molecular axis $\hat L_z$ commutes with the Hamiltonian
\footnote{Due to complete separation of variables one more integral in a form of the second order polynomial in momentum exists \cite{Erikson:1949}, it is closely related to Runge-Lenz vector \cite{Coulson:1967} and commutes with $\hat L_z$; hence, the H$_2^+$ ion in adiabatic (Born-Oppenheimer) approximation is completely-integrable system.}
the eigenstate has a definite magnetic quantum number $\La$. The Hamiltonian is permutationally-symmetric $r_1 \lrar r_2$, or, equivalently, $\eta \rar -\eta$, hence, any eigenfunction is of a definite parity ($\pm$). As a result, it can be represented in a form
\begin{equation}
\label{psi}
   \Psi \ =\ X(\xi) (\xi^2-1)^{\La/2} Y(\eta) (1-\eta^2)^{\La/2} e^{ \pm i \La \phi}\ ,\ \La=0,1,2,\ldots
\end{equation}
where $Y(\eta)$ is of definite parity. After substitution of (\ref{psi}) into (\ref{Sch}) we arrive at the equations for $X(\xi)$ and $Y(\eta)$,
\begin{equation}
\label{X_L}
       \pa_{\xi} [(\xi^2-1) \pa_{\xi} X] + 2 \La \xi \pa_{\xi} X +\left[-p^2\xi^2 + 2R \xi +A\right] X\ =\ 0\ ,\
       X \in L^2(\xi \in [1,\infty))\ ,
\end{equation}
\begin{equation}
\label{Y_L}
    \pa_{\eta} [(\eta^2-1) \pa_{\eta} Y] + 2 \La \eta \pa_{\eta} Y + \left[-p^2\eta^2 + A \right] Y\ =\ 0\ ,\
       Y \in L^2(\eta \in [-1,1])\ ,
\end{equation}
respectively, where following \cite{Bates:1953} we denote,
\begin{equation}
\label{p}
    p^2\ =\ - \frac{E'R^2}{4}\ ,
\end{equation}
and $A$ is a separation constant. Equations (\ref{X_L}), (\ref{Y_L}) define a bispectral problem with $E,A$ as spectral parameters. Both spectral parameters $E,A$ depend on $R$. Square-integrability of the function $\Psi$ (\ref{psi}) implies a non-singular behavior of $X$ at $\xi \rar 1$ as well as non-singular behavior of $Y$ at $\eta \rar \pm 1$. The latter condition implies a certain behavior of the solution $Y$ at large arguments $\eta$. A non-singular solution $Y(\eta)$ can be unambiguously continued in $\eta$ beyond the interval $[-1,1]$, it has to be growing (non-decaying) at $|\eta| \rar \infty$.
%
It is in agreement with large-$\eta$ behavior of the Hund-Mulliken function (it mimics the incoherent interaction of electron with charged centers) for both $1s\si_g$ (parity +) and  $2p\si_u$ (parity -) states
\begin{equation}
\label{HM}
    \Psi^{(\pm)}_{HM}\ =\ e^{- 2\al_2 r_1} \pm e^{- 2\al_2 r_2}\
    =\ 2 e^{-\al_2 R \xi}
    \left[
   \begin{array}{c}
    \cosh (\al_2 R \eta) \\
    \sinh (\al_2 R \eta)
    \end{array}
    \right]
    \ ,
\end{equation}
which describes large $R$ behavior, similarly, for the Guillemin-Zener function (it mimics the coherent interaction of electron with charged centers)
\begin{equation}
\label{GZ}
   \Psi^{(\pm)}_{GZ}\ =\ e^{- 2\al_3 r_1 - 2\al_4 r_2} \pm e^{- 2\al_3 r_2 - 2\al_4 r_1}\
   =\ 2 e^{-(\al_3+\al_4) R \xi}
   \left[
   \begin{array}{c}
    \cosh ((\al_3-\al_4) R \eta) \\
    \sinh ((\al_3-\al_4) R \eta)
    \end{array}
    \right]
    \ ,
\end{equation}
which has to correspond to small $R$ behavior.

\noindent {\it Asymptotics.} If we put $X=e^{-\varphi}$, then at $\xi \rar \infty$,
\begin{equation}
\label{X-inf}
 \varphi\ =\ p\xi - \left( \frac{R}{p}-\La-1 \right) \log \xi + \left[\frac{A+(\frac{R}{p}-\La-1)(\frac{R}{p}+\La)}{p} - p\right]\frac{1}{2\xi}
 +\ldots\ ,
\end{equation}
which is nothing but WKB-expansion, and at $\xi \rar 0$,
\begin{equation}
\label{X-small}
 \varphi\ =\ -\frac{A}{2}\xi^2 - \frac{R}{3}\xi^3 + \frac{(p^2+A^2-A(2\La+3))}{12} \xi^4 + \ldots \ .
\end{equation}
Similarly to $X$ we put $Y = e^{-\varrho}$, then at $\eta \rar \infty$,
\begin{equation}
\label{Y-inf}
 \varrho\ =\ -p\eta + (\La+1)\log \eta - \left(\frac{A-\La(\La+1)}{p} - p\right)\frac{1}{2\eta}
 +\ldots\ ,
\end{equation}
when at $\eta \rar 0$,
\begin{equation}
\label{Y-small}
 \varrho\ =\ -\frac{A}{2}\eta^2 + \frac{(p^2+A^2-A(2\La+3))}{12} \eta^4 + \ldots\ .
\end{equation}
The important property of the expansions (\ref{X-inf}) and (\ref{Y-inf}) is that the coefficients in front of the growing terms at large distances (linear and logarithmic)
are found explicitly, since they do not depend on the separation constant $A$.

\noindent {\it Approximation}.
Making interpolation between WKB-expansion (\ref{X-inf}) and the perturbation theory (\ref{X-small}) for $X$, (\ref{Y-inf}) and (\ref{Y-small}) for $Y$, correspondingly, and taking into account that the $Z_2$-symmetry of $\Psi$: $\eta \rar -\eta$ is realized through use of $\cosh(\sinh)$-function (cf. (\ref{HM}) and (\ref{GZ})) we arrive at the following expression
\begin{equation}
\label{appr}
 \Psi^{(\pm)}_{n,m,\La} = \frac{(\xi^2-1)^{\La/2}P_n (\xi)}{(\gamma + \xi)^{1+\La-\frac{R}{p}}}
 e^{-\xi \frac{\al + p \xi}{\gamma + \xi}}
 \frac{(1-\eta^2)^{\La/2}Q_m(\eta^2)}{(1 + b_2 \eta^2 + b_3 \eta^4)^{\frac{1+\La}{4}}}
 \left[
   \begin{array}{c}
    \cosh  \\
    \sinh
    \end{array}
 \left(\eta \frac{a_1 + p a_2 \eta^2 + p b_3 \eta^4}
 {1 + b_2\eta^2 + b_3 \eta^4}\right) \right] e^{\pm i \La \phi} \ ,
\end{equation}
for the eigenfunction of the state with the quantum numbers $(n,m,\La,\pm)$. Here $\al,\gamma$ and $a_{1,2}, b_{2,3}$ are parameters (see below), $P_n (\xi)$ and $Q_m(\eta^2)$ are some polynomials of degrees $n$ and $m$ with real coefficients with $n$ and $m$ real roots in the intervals $[1,\infty)$ and $[0,1]$, respectively. These polynomials should be chosen in such a way to ensure their orthogonality.

\noindent {\it Results.}
As an illustration we consider two lowest states - one of positive and one of negative parity, $1s \si_g\ (0,0,0,+)$ and $2p \si_u\ (0,0,0,-)$, respectively. Corresponding approximations have the form
\[
 \Psi^{(\pm)}_{0,0,0} = \frac{1}{(\gamma + \xi)^{1-\frac{R}{p}}}
 e^{-\xi \frac{\al + p \xi}{\gamma + \xi}}
 \frac{1}{(1 + b_2 \eta^2 + b_3 \eta^4)^{1/4}}
 \left[
   \begin{array}{c}
    \cosh  \\
    \sinh
    \end{array}
 \left(\eta \frac{a_1 + p a_2 \eta^2 + p b_3 \eta^4}
 {1 + b_2\eta^2 + b_3 \eta^4}\right) \right]
\]
\begin{equation}
\label{appr-0}
   \equiv X_0(\xi) Y_0^{(\pm)}(\eta) \ ,
\end{equation}
(cf.(\ref{appr})) and each of them depends on six parameters $\al,\gamma$ and $a_{1,2}, b_{2,3}$. The easiest way to find these parameters is to make a variational calculation taking (\ref{appr-0}) as a trial function for $R$ fixed and with $p$ as an extra variational parameter. Immediate striking result of the variational study is that for all $R \in [1,50]$ the optimal value of the parameter $p$ coincides with the exact value of $p$ (see (\ref{p})) with extremely high accuracy for both $1s \si_g$ and $2p \si_u$ states. It implies a very high quality of the trial function - the variational optimization wants to reproduce with very high accuracy a domain where the eigenfunction is exponentially small, hence, the domain which gives a very small contribution to the energy functional. In Tables I,II the results for the total energy (as well as for sensitive $p$) vs $R$ of $1s \si_g$ and $2p \si_u$ states are shown as well as their comparison with ones obtained by Montgomery \cite{Montgomery:1977} in highly-accurate realization of the approach by Bates et al \cite{Bates:1953}, and also with the results we obtained in the Lagrange mesh method based on Vincke-Baye approach \cite{Baye:2006} (details will be given elsewhere). For all studied values of $R$ for both $1s \si_g$ and $2p \si_u$ states our variational energy turns out to be in agreement on the level of 10 s.d. with these two alternative calculations. Variational parameters are smooth slow-changing functions of $R$, see Tables III-IV. All calculations were implemented in double precision arithmetics and checked in quadruple precision one. It is worth noting that the number of optimization parameters can be reduced putting $a_2=b_2=0$ - the accuracy in energy drops from 10-11 to 5-6 significant digits.

Hence, our relatively-simple, few parametric functions (\ref{appr-0}) taken as trial functions in a variational study provide very high accuracy in energy. The natural question to ask is how close these functions are to the exact ones in configuration space.
In order to study this question we develop a perturbation theory in the Schroedinger equation (\ref{Sch}) taking a trial function (\ref{appr-0}) as zero approximation. The easiest way to realize it is to consider non-linearization procedure \cite{Turbiner:1984}: if the potential is of the form $V=V_0 + \la V_1$, then it is looked for energy and the eigenfunction in the form of power-like series in the parameter $\la$, $E=\sum \la^n E_n$ and $\Psi=\Psi_0 \exp (-\sum \la^n \varphi_n)$, respectively.
Due to specifics of (1) because of the separation of variables the procedure can be developed for both functions $X$ and $Y$ (see (\ref{psi})) separately as well as for the separation parameter $A$, while keeping the energy $E$ fixed. It can be done for the system of equations (\ref{X_L}), (\ref{Y_L}). For simplicity we consider nodeless in $\xi$ and $\eta$ states, $(0,0,\La,\pm)$. As a first step let us transform (\ref{X_L}), (\ref{Y_L}) into the Riccati form by introducing $X=e^{-\varphi}$ and $Y=e^{-\varrho}$, respectively,
\begin{equation}
\label{X_L-phi}
       (\xi^2-1)(x' - x^2) + 2 (\La+1) \xi\ x\ =\
       A - V(\xi) \ ,\quad x=\varphi'_{\xi}
\end{equation}
where the ``potential" $V(\xi) = a^2\xi^2 - 2R \xi$, and
\begin{equation}
\label{Y_L-rho}
       (\eta^2-1)(y' - y^2) + 2 (\La+1) \eta\ y\ =\
       A - W(\eta) \ ,\quad y = \varrho'_{\eta}
\end{equation}
where the ``potential" $W(\eta) = a^2\eta^2$.

Let us choose some $x_0(\xi)=\varphi_0'(\xi)$, then substitute it to the l.h.s. of (\ref{X_L-phi}) and call the result as unperturbed "potential" $V_0(\xi)$ putting without loss of generality $A_0=0$. The difference between the original $V(\xi)$ and generated $V_0(\xi)$ is the perturbation, $V_1(\xi)=V(\xi)-V_0(\xi)$. For a sake of convenience we can insert a parameter $\la$ in front of $V_1$ and develop the perturbation theory in powers of it,
\begin{equation}
\label{PTx}
    x=\sum \la^n x_n\ , \ A=\sum \la^n A_{n,x}\ .
\end{equation}
The equation for $n$th correction has a form,
\begin{equation}
\label{x_n}
    (\xi^2-1)x_n' - 2[(\xi^2-1)x_0 - (\La+1)\xi]x_n\ =\ A_{n,x} - v_n
\end{equation}
where $v_1=V_1$ and $v_n= (\xi^2-1)\sum_{i=1}^{n-1} x_i x_{n-i}$ for $n>1$. It can be immediately solved,
\begin{equation}
\label{x_n-solu}
    (\xi^2-1)^{\La+1}e^{-2 \varphi_0} x_n\ =\ \int_1^{\xi} (A_{n,x} - v_n) (\xi^2-1)^{\La}e^{-2 \varphi_0} d\xi\ ,
\end{equation}
and
\begin{equation}
\label{xA_n}
    A_{n,x}\ =\ \frac{\int_1^{\infty} v_n (\xi^2-1)^{\La}e^{-2 \varphi_0} d\xi}
    {\int_1^{\infty} (\xi^2-1)^{\La}e^{-2 \varphi_0} d\xi}
\end{equation}
In a similar manner by choosing $y_0(\eta)=\varrho_0'(\eta)$, building the unperturbed "potential" $W_0(\eta)$ and putting $A_0=0$ as zero approximation one can develop perturbation theory in the equation (\ref{Y_L-rho})
\begin{equation}
\label{PTy}
    y=\sum \la^n y_n\ , \ A=\sum \la^n A_{n,y}\ .
\end{equation}
The equation for $n$th correction has a form similar to (\ref{x_n}),
\begin{equation}
\label{y_n}
    (\eta^2-1)y_n' - 2[(\eta^2-1)y_0 - (\La+1)\eta]y_n\ =\ A_{n,y} - w_n
\end{equation}
where $w_1=W_1\equiv W-W_0$ and $w_n= (\eta^2-1)\sum_{i=1}^{n-1} y_i y_{n-i}$ for $n>1$. Its solution is given by
\begin{equation}
\label{y_n-solu}
    (\eta^2-1)^{\La+1}e^{-2 \varrho_0} y_n\ =\ \int_{-1}^{\eta} (A_{n,y} - w_n) (\eta^2-1)^{\La}e^{-2 \varrho_0} d\eta\ ,
\end{equation}
(cf.(\ref{x_n-solu})) and
\begin{equation}
\label{yA_n}
    A_{n,y}\ =\ \frac{\int_{-1}^{1} w_n (\eta^2-1)^{\La}e^{-2 \varrho_0} d\eta}
    {\int_{-1}^{1} (\eta^2-1)^{\La}e^{-2 \varrho_0} d\eta}
\end{equation}
(cf.(\ref{xA_n})). In order to realize this perturbation theory a condition of consistency should be imposed
\begin{equation}
\label{An}
    A_{n,x}\ =\ A_{n,y}\ .
\end{equation}
This condition allows us to find the parameter $p$ and, hence, the energy $E'$ and $E$ (see (\ref{p})).

Sufficient condition for such a perturbation theory to be convergent is to require a perturbation "potential" to be bounded,
\begin{equation}
\label{PerPot}
    |V_1(\xi)| \leq C_{\xi}\ ,\ |W_1(\eta)| \leq C_{\eta}\ ,
\end{equation}
where $C_{\xi}, C_{\eta}$ are constants. Obviously, that the rate of convergence gets faster with smaller values of $C_{\xi}, C_{\eta}$. It is evident that the perturbations $V_1(\xi)$ and $W_1(\eta)$ get bounded if $\varphi_0(\xi)$ and $\varrho_0(\eta)$ are smooth functions vanishing at the origin but reproduce exactly the growing terms at $|\xi|, |\eta|$ tending to infinity in (\ref{X-inf}), (\ref{Y-inf}), respectively.

Let us choose $X_0, Y_0$ (\ref{appr-0}) with parameters fixed variationally (see above) as zero approximation in perturbation theory (\ref{PTx}), (\ref{PTy}). By construction of $X_0, Y_0$ the emerging perturbation theory is convergent. Assuming the condition (\ref{An}) fulfilled for the first corrections, namely, $A_{1,x}\ =\ A_{1,y}=A_1$, we find the first corrections $\varphi_1(\xi)$ and $\varrho_1(\eta)$ as functions of $A_1$. Then we modify the trial function (\ref{appr-0}),
\begin{equation}
\label{appr-1}
\Psi^{(\pm)}_{0,0,0} \rar X_0(\xi) Y_0^{(\pm)}(\eta)\ e^{-\varphi_1(\xi)-\varrho_1(\eta)}
\end{equation}
and make the variational calculation with this trial function minimizing with respect to parameter $p$. The result is that the optimal value of parameter $p$ remained unchanged with respect to the value obtained for the trial function (\ref{appr-0}) with extremely high accuracy - within 10 s.d.! It indicates that the condition (\ref{An}) is fulfilled with high accuracy. The variational energy is changed beyond the 10 s.d. Therefore, the our energies presented in Tables I,II are correct in all digits. The separation parameters $A_{1,\xi}, A_{1,\eta}$ are presented in Table V. It allows us to find explicitly $\varphi_1(\xi)$ and $\varrho_1(\eta)$. As an illustration in Figs. 1-4 the functions $X_0(\xi) Y_0^{(\pm)}(\eta)$ and the first corrections to them are shown for $R=2$ a.u.

\begin{table}[ht]
\begin{center}
\caption{The total energy $E_t(R)$ for $1s \si_g$ state of the H$_2^+$-ion compared to \cite{Montgomery:1977} (rounded) and Lagrange mesh method.}
\begin{tabular}{ccc}
\hline \hline
            R[a.u.] & $E_t$[Ry] (Present/\cite{Montgomery:1977}/Mesh)
            &\qquad $p$ \hspace{0.9cm} \phantom{.}\\
\hline \hline
\multirow{3}{*}{1.0}      &-0.90357262676  &\,\,0.8519936\\
                          &-0.90357262676  & \\
                          &-0.90357262676  & \\
\br
\multirow{3}{*}{1.997193}  & -1.20526923821   &\,\,1.483403 \\
                           & \qquad --        & \\
                           & -1.20526923821   &\\
\br
\multirow{3}{*}{2.0}      &-1.20526842899  &\,\,1.485015\\
                          &-1.20526842899  & \\
                          &-1.20526842899  & \\
\br
\multirow{3}{*}{6.0}      &-1.0239380968   &\,\,3.49506\\
                          &-1.0239380969   &\\
                          &-1.0239380969   &\\
\br
\multirow{3}{*}{10.0}     &-1.0011574578   &\,\,5.47987\\
                          &-1.0011574579   &\\
                          &-1.0011574579   &\\
\br
\multirow{3}{*}{12.5}     &-1.0002611115  &\,\,6.73221\\
                          &\,\,\,-----    &\\
                          &-1.0002611116  &\\
\br
\multirow{3}{*}{30.0}     &-1.0000055815   & 15.492\\
                          &\,\,\,-----     &\\
                          &-1.0000055815   &\\
\br
\multirow{3}{*}{40.0}     &-1.0000017622   & 20.4939\\
                          &\,\,\,-----     &\\
                          &-1.0000017622   &\\
\br
\multirow{3}{*}{50.0}     &-1.0000007211   & 25.49511\\
                          &\,\,\,-----     &\\
                          &-1.0000007211   &\\
\hline \hline
\end{tabular}
\end{center}
\end{table}

\begin{table}[ht]
\begin{center}
\caption{The total energy $E_t(R)$ for $2p \si_u$ state of the H$_2^+$-ion compared to \cite{Montgomery:1977} (rounded) and Lagrange mesh method.}
\begin{tabular}{ccc}
\hline \hline
            R\ [a.u.] &\phantom{.}\, $E_t$\ (Present/\cite{Montgomery:1977}/Mesh)\ [Ry]
            &\phantom{.} \qquad $p$ \hspace{0.9cm} \phantom{.}\\
\hline \hline
\multirow{3}{*}{1.0}   &\,\,0.8703727499   & 0.5314196\\
                       &\,\,0.8703727498   & \\
                       &\,\,0.8703727498   & \\
\br
\multirow{3}{*}{1.997193} &-0.3332800331   & 1.1536645\\
                          &\,\,\,-----     & \\
                          &-0.33328003316  &\\
\br
\multirow{3}{*}{2.0}      &-0.3350687844   & 1.155452\\
                          &-0.3350687844   &\\
                          &-0.3350687844   &\\
\br
\multirow{3}{*}{4.0}      &-0.8911012787   & 2.3589\\
                          &-0.8911012787   &\\
                          &-0.8911012787   &\\
\br
\multirow{3}{*}{10.0}     &-0.9998021372   &5.47678\\
                          &-0.9998021372   &\\
                          &-0.9998021372   &\\
\br
\multirow{3}{*}{12.54525}  &-1.0001215811   & 6.75434 \\
                          &\,\,\,---       & \\
                          &-1.0001215811   &\\
\br
\multirow{3}{*}{20.0}     &-1.0000283953   & 10.4882\\
                          &-1.0000283953   &\\
                          &-1.0000283953   &\\
\br
\multirow{3}{*}{30.0}     &-1.0000055815   & 15.492\\
                          &\,\,\,---        &\\
                          &-1.0000055815   &\\
\br
\multirow{3}{*}{40.0}     &-1.0000017622   & 20.4939 \\
                          &\,\,\,---        &\\
                          &-1.0000017622   &\\

\hline \hline
\end{tabular}
\end{center}
\end{table}

\begin{table}[ht]
\caption{The parameters of the function (\ref{appr-0})
for $1s \si_g$ state of the H$_2^+$-ion. The parameters $\al,\gamma$ and $a_{1,2}, b_{2,3}$ are found via minimization.}
\begin{center}
\begin{tabular}{c|c|c|c}
\hline
\hline
 &\ $R_{eq}$=1.997193 a.u. &\ $R$=6.0 a.u. &\ $R$=20.0 a.u.\\
\hline\hline
\br
   $\al$\  & 1.48407      & 3.32381      & 10.0453     \\
   $p$\    & 1.483403     & 3.49506      & 10.4882     \\
  $\gamma$ & 1.0299       & 0.96357      & 0.95774    \\
\br
$a_1$  & 0.9164     & 2.597355   & 9.8775      \\
$a_2$  & 0.05384    & 0.53443    & 6.8392     \\
$b_2$  & 0.06       & 0.588072   & 6.9016     \\
$b_3$  & 0.00011    & 0.00552    & 1.352 \\
\hline
\hline
\end{tabular}
\end{center}
\end{table}

\begin{table}[ht]
\caption{The parameters of the function (\ref{appr-0})
for $2p \si_u$ state of the H$_2^+$-ion. The parameters $\al,\gamma$ and $a_{1,2}, b_{2,3}$ are found via minimization.}
\begin{center}
\begin{tabular}{c|c|c|c}
\hline
\hline
 & $R$=6.0 a.u. &\ $R_{min}=$12.54525 a.u. & R=20.0 a.u.\\
\hline
\hline
\br
 $\al$     & 3.24715      & 6.5275       & 10.7397     \\
 $p$       & 3.43971      & 6.75434      & 10.4882      \\
 $\gamma$  & 0.95706      & 0.97045      & 1.03027    \\
\br
$a_1$  & 2.84566    & 6.075      & 9.8077      \\
$a_2$  & 0.22098    & 1.46757    & 2.3784     \\
$b_2$  & 0.23611    & 1.5349     & 2.43705    \\
$b_3$  &-0.0027     & 0.1675      & 0.367       \\
\hline
\hline
\end{tabular}
\end{center}
\end{table}

\begin{table}[h]
\caption{Separation parameters $A_{1,\xi}, A_{1,\eta}$  for $1s \si_g, 2p \si_u$ states of the H$_2^+$-ion compared to Scott et al \cite{Scott:2006}.}
\begin{center}
\begin{tabular}{c|ccc|ccc}
\hline \hline
       &\multicolumn{3}{c}{$1s\si_g$}\vline&\multicolumn{3}{c}{$2p\si_u$}\\
\cline{2-4} \cline{5-7}
  $R$  &  $A_{1,\xi}$      & $A_{1,\eta}$  &\qquad \cite{Scott:2006}
       &  $A_{1,\xi}$      & $A_{1,\eta}$  &\qquad \cite{Scott:2006}
  \\
\hline\hline
 2.0   &   0.811729588 &   0.811729585 & 0.811729585
       &  -1.186889395 &  -1.186889393 & -1.18688939 \\
15.0   &  48.822353534 &  48.822353528 & ---
       &  48.821470973 &  48.821470957 & --- \\
20.0   &  90.052891187 &  90.052891183 & 90.0528912
       &  90.052877564 &  90.052877564 & 90.0528776 \\
30.0   & 210.034596601 & 210.034596599 & ---
       & 210.034596601 & 210.034596599 & --- \\
\hline
\end{tabular}
\end{center}
\end{table}

Knowledge of wave functions with high local relative accuracy gives us a chance to calculate matrix elements with controlled relative accuracy $\lesssim 10^{-5}$. As a demonstration we calculate the Oscillator Strength as function of interproton distance for the simplest radiative transition $2p \si_u \rar 1s\si_g$ are (see e.g. \cite{Bishop:1978}),
\begin{equation}
\label{OS}
f_{01}(R) = \frac{2}{3}(E^{2p \si_u}(R)-E^{1s\si_g}(R))|{\bf Q}(R)|^2\ ,
\end{equation}
where ${\bf Q}(R)$ is the matrix element
\[
 {\bf Q}(R)=\langle \Psi^{1s\si_g}(R)| {\bf r} |\Psi^{2p \si_u}(R)\rangle\ ,
\]
where ${\bf r}$ is the vector of the electron position measured from the internuclear midpoint, and wavefunctions $\Psi^{1s\si_g}, \Psi^{2p \si_u}$ are given by (\ref{appr-0}). It is assumed this calculation should provide at least 5 s.d. correctly. In Table VI the results are presented. For all internuclear distances they coincide in 2 s.d. with Bishop et al \cite{Bishop:1978}, thus, indicating the 3rd digit obtained in \cite{Bishop:1978} is incorrect for $R=1,2,4$ a.u., and in 6 figures with recent results \cite{ts:2010} (with an exception at $R$=1 a.u. where it deviates in one unit at the 6th digit) which increases up to 8 figures for large $R$. Modification of (\ref{appr-0}) by adding the first corrections (\ref{appr-1}) and use it in (\ref{OS}) does not change our 6 s.d. in Table VI.

Summarizing we want to state that a simple uniform approximation of the eigenfunctions for the H$_2^+$ molecular ion is presented. It allows us to calculate any expectation value or matrix element with guaranteed accuracy. It manifests the approximate solution of the problem of spectra of the H$_2^+$ molecular ion. In a quite straightforward way similar approximations can be constructed for general two-center, one-electron system $(Z_a,Z_b,e)$, in particular, for $(\rm HeH)^{++}$. It will be done elsewhere.

The key element of the procedure is to construct an interpolation between the WKB expansion at large distances and perturbation series at small distances for the phase of the wavefunction. Or, in other words, to find an approximate solution for the corresponding eikonal equation. Separation of variables allowed us to solve this problem. In the case of non-separability of variables the WKB expansion of a solution of the eikonal equation can not be constructed in unified way, since all depends on the way to approach to infinity. However, a reasonable approximation of the first growing terms of the WKB expansion seems sufficient to construct the interpolation between large and small distances giving high accuracy results. This program was realized for the problem of the hydrogen atom in a magnetic field and will be published elsewhere.

It is worth mentioning a curious fact that the problem (\ref{Sch}) possesses the hidden algebra $sl(2)\oplus sl(2)$.
It can be immediately seen - making the gauge rotation of the operators in r.h.s. of the equations (\ref{X_L}) and (\ref{Y_L}) with gauge factors $e^{-p\xi}$ and  $e^{p\eta}$, respectively. We obtain the operators which are in the universal enveloping algebra of $sl(2)$ (see e.g. \cite{Turbiner:1988}). The dimension of the representation is $-\La$ and $-\La+\frac{R}{p}$, respectively. For non-physical values of $\La$ and integer ratio $\frac{R}{p}$ the algebras $sl(2)$ appear in the finite-dimensional representation realized in action on polynomials in $\xi, \eta$. It explains a mystery sometimes observed of the existence of polynomial solutions for non-physical values of $\La$ in the problem (2) (details will be given elsewhere).

\textit{\small Acknowledgements}.  The research is supported in part by DGAPA grant IN115709 and CONACyT grant 58942-F (Mexico). H.O.P. is supported by CONACyT project for postdoctoral research. A.V.T. thanks the University Program FENOMEC (UNAM, Mexico) for partial support.

\begin{table}[h]
\caption{Oscillator strength $f_{01}$ (\ref{OS}) for transition $2p \si_u \rar 1s\si_g$ vs $R$ compared to Bishop et al \cite{Bishop:1978} and Tsogbayar et al \cite{ts:2010} (rounded).}
\begin{center}
\begin{tabular}{cccc}
\hline \hline
R        &  present       & \cite{Bishop:1978} &\qquad \cite{ts:2010} \\
\hline\hline
1.0      & 0.538675             & \ 0.538\  & \ 0.5386739 \\
1.997193 & 0.639595             & \quad ---   & \quad ---  \\
2.0      & 0.639527             & \ 0.638\  & \ 0.6395268 \\
4.0      & 0.469200             & \ 0.476\  & \ 0.4692004 \\
10.0     & 2.217 $\times10^{-02}$ & 0.022\  & \ 2.21706 $\times 10^{-02}$ \\
15.0     & 5.129 $\times10^{-04}$ &  ---    & \ 5.12939 $\times 10^{-04}$ \\
20.0     & 8.191 $\times10^{-06}$ &  ---    & \ 8.20513 $\times 10^{-06}$ \\
30.0     & 4.770 $\times10^{-09}$ &  ---    & \quad --- \\
40.0     & 1.828 $\times10^{-10}$ &  ---    & \quad --- \\
\hline \hline
\end{tabular}
\end{center}
\end{table}
\begin{center}
\begin{figure}
\begin{tabular}{cc}
\subfloat[]{
\includegraphics[width=3in,height=3.2in,angle=0]{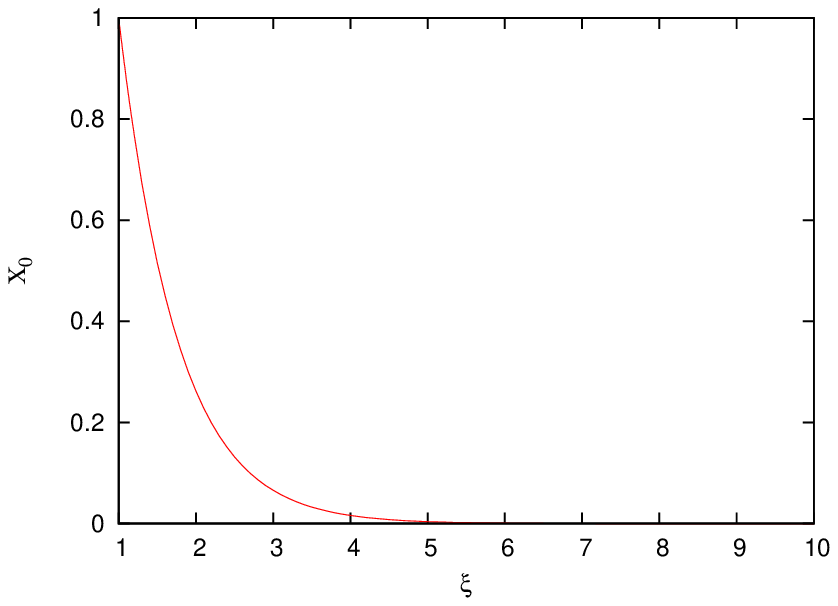}}
&
\subfloat[]{
\includegraphics[width=3in,height=3.2in,angle=0]{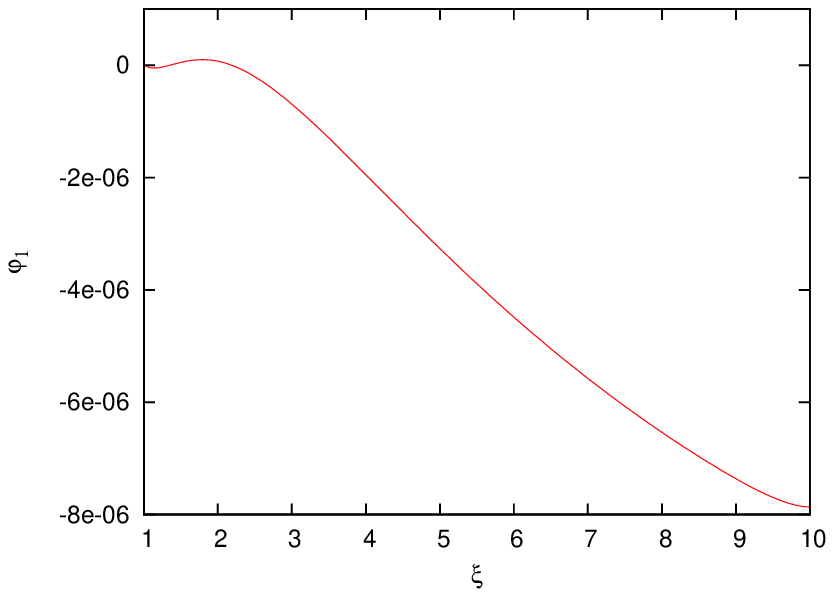}}
\end{tabular}
\caption{\label{fig-1sg-X}
The $1s\si_g$ state  at $R=2$ a.u.: (a)\ $\xi$-dependent function $X_0$ (\ref{appr-0}) and (b)\ the first correction $\phi_1$ (see (\ref{appr-1})).}
\end{figure}
\end{center}
\begin{center}
\begin{figure}
\begin{tabular}{cc}
\subfloat[]{
\includegraphics[width=3in,height=3.2in,angle=0]{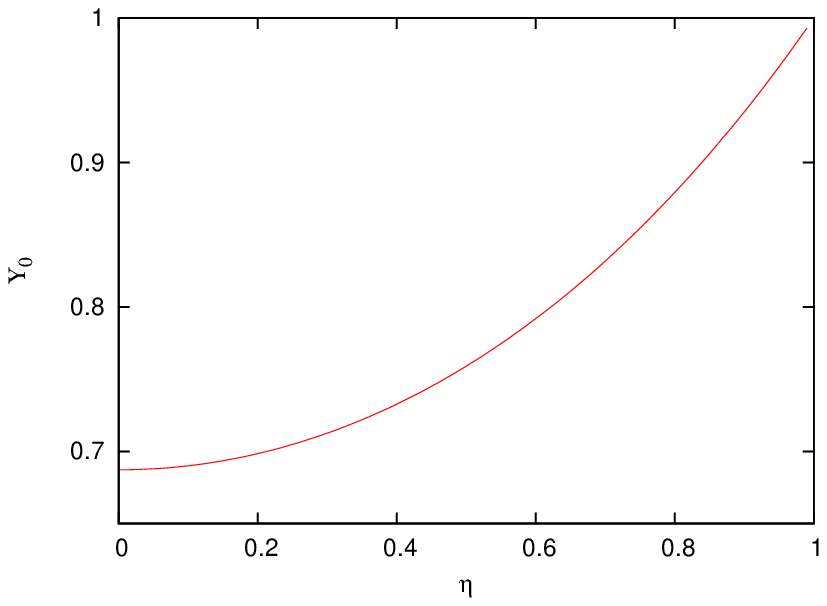}}
&
\subfloat[]{
\includegraphics[width=3in,height=3.2in,angle=0]{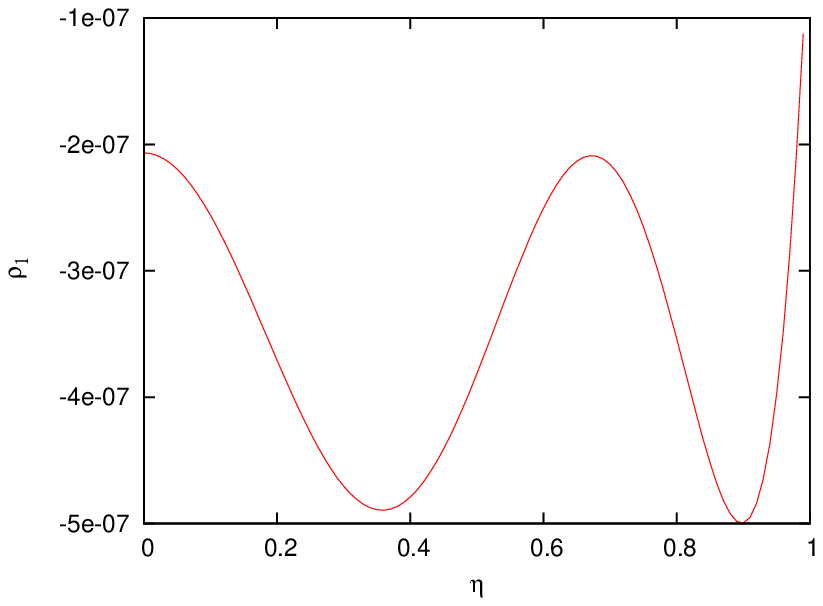}}
\end{tabular}
\caption{\label{fig-1sg-Y}
The $1s\si_g$ state at $R=2$ a.u.: (a)\ $\eta$-dependent function $Y_0^{(+)}$ (\ref{appr-0}) and (b)\ the first correction $\rho_1$ (see (\ref{appr-1})).}
\end{figure}
\end{center}
\begin{center}
\begin{figure}
\begin{tabular}{cc}
\subfloat[]{
\includegraphics[width=3in,height=3.2in,angle=0]{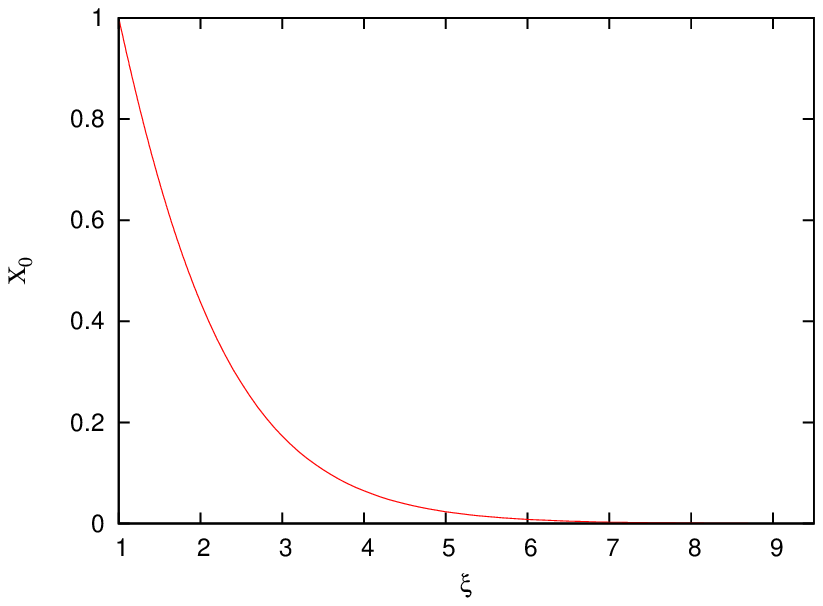}}
&
\subfloat[]{
\includegraphics[width=3in,height=3.2in,angle=0]{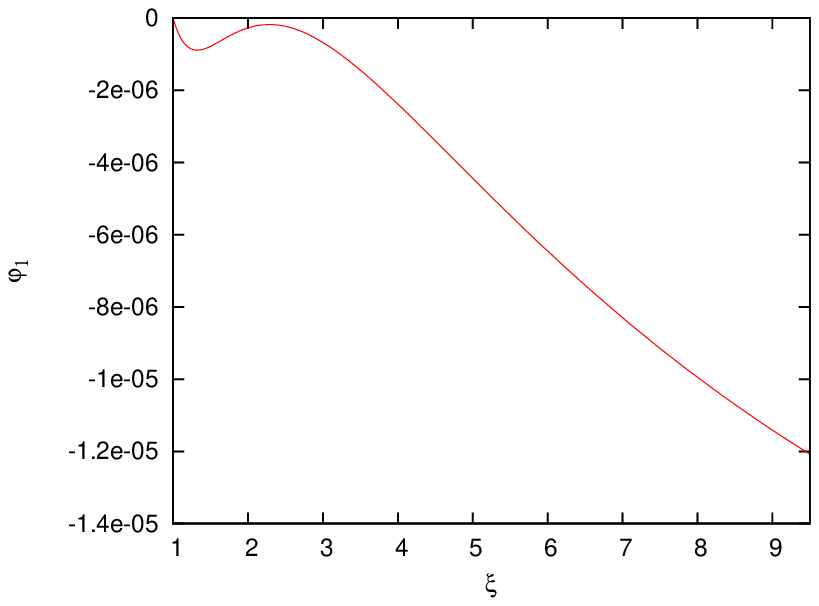}}
\end{tabular}
\caption{\label{fig-1su-X}
The $2p\si_u$ state at $R=2$ a.u.: (a)\ $\xi$-dependent function $X_0$ (\ref{appr-0}) and (b)\ the first correction $\phi_1$ (see (\ref{appr-1})).}
\end{figure}
\end{center}
\begin{center}
\begin{figure}
\begin{tabular}{cc}
\subfloat[]{
\includegraphics[width=3in,height=3in,angle=0]{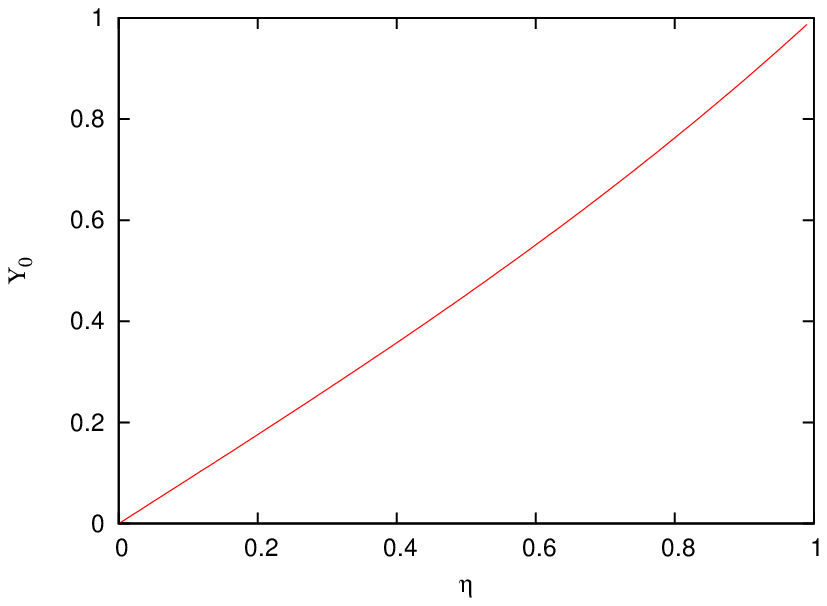}}
&
\subfloat[]{
\includegraphics[width=3in,height=3in,angle=0]{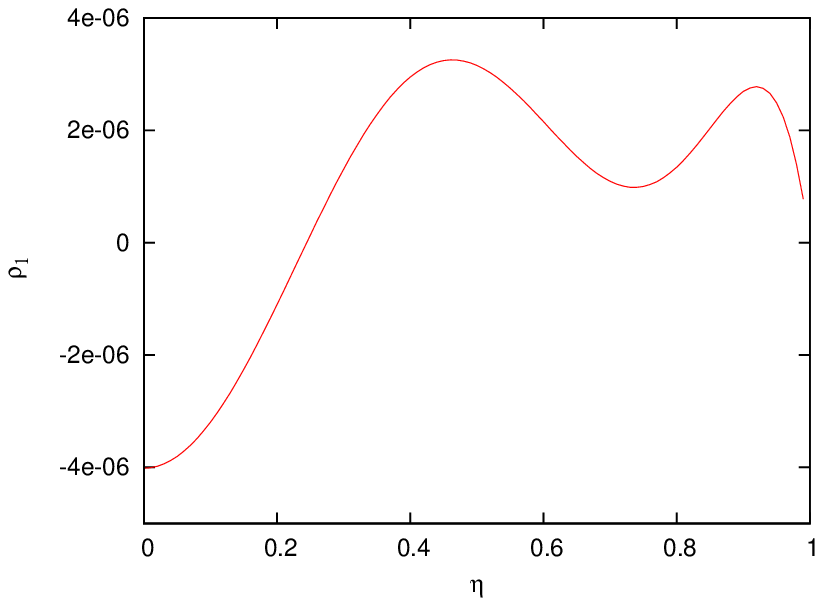}}
\end{tabular}
\caption{\label{fig-1su-Y}
The $2p\si_u$ state at $R=2$ a.u.: (a)\ $\eta$-dependent function $Y_0^{(-)}$ (\ref{appr-0}) and (b)\ the first correction $\rho_1$ (see (\ref{appr-1})).}
\end{figure}
\end{center}

\end{document}